\documentstyle[twocolumn,prl,floats,aps,psfig]{revtex}
\newcommand{\mywidth}{3.8in}

\begin{document}
\draft
\twocolumn[\hsize\textwidth\columnwidth\hsize\csname
@twocolumnfalse\endcsname

\title{Scaling of the superfluid density in superfluid films}   
\author{Norbert Schultka$^1$ and Efstratios Manousakis$^2$}
\address{$^1$ Institut f\"ur Theoretische Physik, Technische
Hochschule Aachen, D--52056 Aachen, Germany\\ $^2$Department of Physics
and Center for Materials Research and Technology, \\
Florida State University, Tallahassee, Florida 32306, USA}
\date{\today}
\maketitle
\begin{abstract}
We study scaling of the superfluid density with respect to the film 
thickness by simulating the $x-y$ model on films of 
size $L \times L \times H$ ($L >> H$) using the  cluster Monte Carlo.
While periodic boundary conditions where used in the planar ($L$)
directions, Dirichlet boundary conditions where used along the film
thickness. We find that our results can be scaled on a universal curve
by introducing an effective thickness. In the limit of large $H$ our
scaling relations reduce to the conventional scaling forms.
Using the same idea we find scaling in the
experimental results using the same value of $\nu = 0.6705$.
\end{abstract}
]
Scaling is a central idea in critical phenomena near a second order
phase transition and in field theory when we are interested in the 
continuum limit\cite{fss}. In both cases we are looking at the 
singular behavior emerging from the 
overwhelmingly large number of degrees of freedom, corresponding to 
the original cutoff scale, which need to be integrated out leaving 
behind long-wavelength degrees of freedom which vary smoothly. 
Their behavior is controlled by a dynamically generated 
length scale, the correlation length $\xi$. Such a fundamental idea 
is difficult to test theoretically because it
requires a study of an overwhelmingly large number of 
interacting degrees of freedom. Experimentally, however, one hopes to be  
able to study scaling in finite-size real systems near a second
order phase transition. Namely, the system is confined in a finite
geometry (for example, film geometry) and the finite-size scaling theory 
is expected to describe the behavior of the system near the bulk critical
temperature $T_{\lambda}$. Liquid $^4He$ has been a good real system
for testing finite-size scaling theory and measuring  the critical
exponents that are associated with the most singular behavior in 
thermodynamic quantities near $T_{\lambda}$. However, measurements of the
superfluid density\cite{rhee} and the specific heat\cite{earlyc} on
helium films fail to verify the finite-size scaling theory. 

The situation of the specific heat has been recently 
clarified\cite{usprl,WACKERDOHM,dohmstau} where the choice of the 
boundary conditions
was a key factor in comparing the more recent measurements of the
universal function\cite{lipa} and that obtained theoretically. 
On the other hand, while new experiments for the specific heat under
confined geometries have been planned to be conducted under more ideal
microgravity conditions\cite{lipa2}, the problems related to the
measurements of the superfluid density\cite{rhee} are still outstanding. 

In this paper we use the $x-y$ model and  the cluster Monte Carlo
method to calculate the {\it superfluid density} on films of size $L \times L
\times H$ ($L >> H$) with periodic boundary conditions in the planar
$L$-directions and Dirichlet boundary conditions (vanishing order
parameter) along the film
thickness dimension.  The same model, geometry, and boundary conditions
where used in Ref.\cite{usprl} to calculate the {\it specific heat}, and
a very good agreement between the theoretically calculated and
experimentally determined universal functions was found. 
In this paper we show that the superfluid density is a far
more sensitive observable than the specific heat with respect to the
requirement that one needs to use very thick films ($H \to \infty$) to
verify scaling with respect to the film thickness $H$.
We have found that
in order to achieve scaling for rather small values of $H$ (as in the
case of the specific heat) we need to modify the scaling expressions by
using a concept of an effective thickness. Our introduction of an effective
thickness $H_{eff} = H + D$ (where $D$ is a finite dynamically
generated length scale) 
is no violation of scaling, since at large $H_{eff}$ the constant $D$ 
can be neglected. Scaling for all the values of $H$ used in
our calculation is achieved with the expected value of 
$\nu=0.6705$\cite{golahl}.
A similar modification to the scaling formula allows scaling
of the experimental results of Rhee et al.\cite{rhee} for the superfluid
density with the same value of $\nu$. 

For the $x-y$ model on a lattice, the helicity modulus
$\Upsilon_\mu(T)/J$ as 
defined in Refs. \cite{teiya} is calculated as 
the ensemble average of
$1/V \bigl (\sum_{\langle i,j \rangle} \cos(\theta_{i}-\theta_{j})(
\vec{e}_{\mu} \cdot \vec{\epsilon}_{ij} )^{2} - 
\beta (\sum_{\langle i,j \rangle} \sin(\theta_{i}-\theta_{j}) \vec{e}_{\mu}
   \cdot \vec{\epsilon}_{ij} )^{2}\bigr )$.
Here $V$ is the volume of the lattice, $\beta=J/k_{B}T$, 
$\vec{e}_{\mu}$ is the unit vector in the corresponding bond direction, and 
$\vec{\epsilon}_{ij}$ is the vector connecting the lattice sites $i$ and $j$.
In the following we omit the vector index since we will always refer to
the $x$-component of the helicity modulus and due to the isotropy
$\Upsilon_{x}=\Upsilon_{y}$. The connection between the
helicity modulus and the superfluid density $\rho_{s}$ is established by  
the relation\cite{fibaja}
$\rho_s(T) = ({m}/{\hbar})^{2} \Upsilon(T)$
where $m$ denotes the mass of the helium atom.

In Ref. \cite{us1} we studied the helicity modulus 
$\Upsilon$ for the $x-y$ model in a film geometry with {\it periodic} boundary 
conditions in the $H$--direction. In a certain temperature range around 
the bulk critical temperature $T_{\lambda}$ where the bulk correlation 
length $\xi(T)$ becomes of the order of the film thickness $H$
the quantity $\Upsilon H/T$ exhibits effectively two--dimensional behavior and 
a Kosterlitz--Thouless phase transition takes place at a temperature 
$T_{c}^{2D}(H) < T_{\lambda}$. We found that
the critical temperature $T_{c}^{2D}(H)$ approaches
$T_{\lambda}$ in the limit $H \rightarrow \infty$ as 
\begin{equation}
  T_{c}^{2D}(H)=T_{\lambda}\left( 1+\frac{x_{c}}{H^{1/\nu}} \right),
 \label{tch}
\end{equation}
where the critical exponent $\nu$ is the same as
the experimental value $\nu=0.6705$ \cite{golahl} and 
the value $T_{\lambda}/J=2.2017$ \cite{jantl}.
We also demonstrated that $\Upsilon H/T$ is a function of the ratio 
$H/\xi(T)$, i.e. the dimensionless quantity
\begin{equation}
  \frac{\Upsilon(T,H) H}{T} = \Phi(tH^{1/\nu}),
  \label{tyh}
\end{equation}                  
is a function of $x=tH^{1/\nu}$ only. We found that when we plotted the 
calculated $\Upsilon(T,H)H/T$ as a function of $x$, in the limit 
$L \rightarrow \infty$, our
results for all thicknesses $H$ collapse on the same universal
curve. Thus, simple scaling holds for periodic boundary conditions.

\begin{figure}[htp] 
 \centerline{\psfig {figure=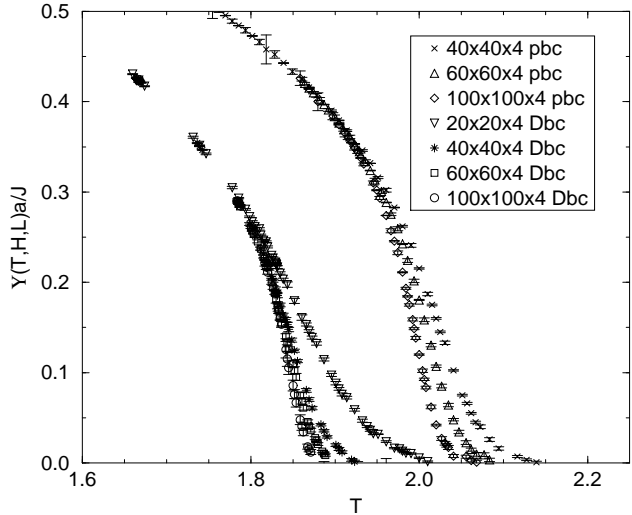,width=\mywidth}}
 \caption{\label{fig1}  The helicity modulus $\Upsilon(T,H,L)$ as
 a function of $T$ for various lattices $L^{2} \times 4$ with Dirichlet
 boundary conditions (Dbc) in the $H$--direction.}
\end{figure}

In this paper we consider periodic boudary conditions in the planar 
$L$-directions
and Dirichlet boundary conditions along the thickness direction.
Fig.\ref{fig1} displays our Monte Carlo data for the helicity modulus
in units of the lattice spacing $a$ and the energy scale $J$ for the 
film of fixed thickness $H=4$. Dirichlet boundary conditions 
strongly suppress the values of the helicity modulus as compared to the
case of periodic boundary conditions along $H$. As a consequence, films with
Dirichlet boundary conditions have lower critical temperatures
than films with periodic boundary conditions.

We eliminate the finite size effects in  the $L$-direction by
studying the quantity 
$K=T/(\Upsilon H)$ in the $L \rightarrow \infty$ limit.
For a fixed thickness $H$ and 
at temperatures $T$ below but sufficiently close to the critical 
temperature $T_{c}^{2D}(H)$,
the system behaves effectively two--dimensionally \cite{us1,AHNS}. It
was demonstrated in Ref. \cite{us1} that we can use the 
Kosterlitz--Thouless--Nelson renormalization group equations \cite{nelko}
to derive an expression for the planar $L$-dependence of $K$. This 
expression can be used to extrapolate the computed values $K(T,H,L)$ 
obtained on lattices of finite $L$ to the $L=\infty$ limit, for a fixed $H$. 
This has been clearly demonstrated in Ref. \cite{us1} for the case 
of periodic boundary conditions. The quality of our extrapolation
is the same as in Ref. \cite{us1} and we omit such demonstration here
due to lack of space. In the following we shall drop the dependence of 
$\Upsilon$ on $L$ implying that we refer to the extrapolated $L \to
\infty$ values.

\begin{figure}[htp] 
 \centerline{\psfig {figure=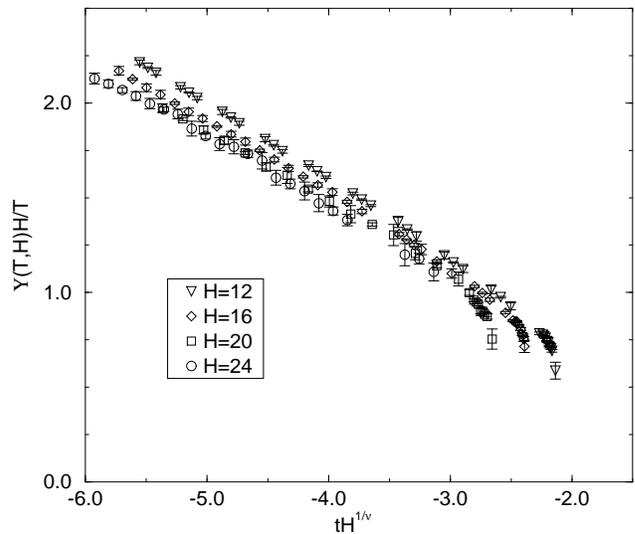,width=\mywidth}}
 \caption{\label{fig2}  $\Upsilon(T,H) H/T$ as a function of $tH^{1/\nu}$
 for various thicknesses. $\nu=0.6705$.}
\end{figure}

In Fig.\ref{fig2} we plot $\Upsilon(T,H)H/T$ versus $tH^{1/\nu}$ for
the thicknesses $H=12,16,20,24$ to check the validity of 
the scaling form (\ref{tyh}) using the experimental value of 
$\nu=0.6705$\cite{golahl}. We do not obtain a universal 
scaling curve, thus scaling according to the expression (\ref{tyh}) is 
not valid for the films with thicknesses up to $H=24$. 

\begin{figure}[htp] 
 \centerline{\psfig {figure=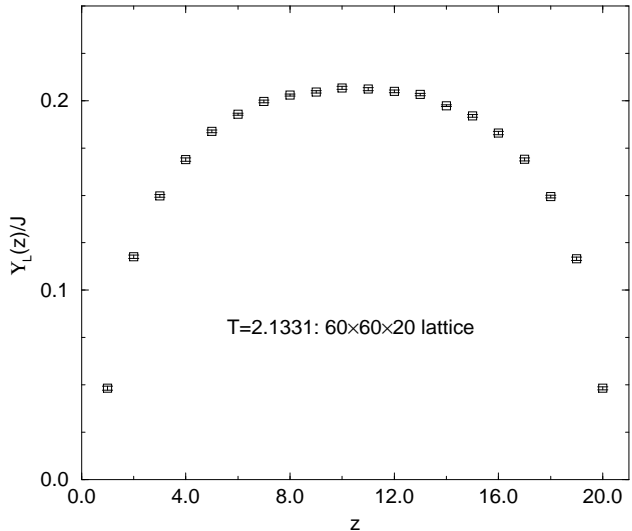,width=\mywidth}}
 \caption{\label{fig3}  The approximate profile $\Upsilon_{L}(z)$ of the
 helicity modulus computed on a $60 \times 60 \times 20$ lattice at
 $T=2.1331$, i.e. close to the critical temperature, $T_{c}^{2D}(20)=2.1346$.}
\end{figure}

Let us, therefore, pursue another line of thought.
In Fig.\ref{fig3} we show the layered helicity modulus 
$\Upsilon_{L}(z)/J$,where $z$ counts the layers, computed on a 
$60 \times 60 \times 20$ lattice at the
temperature $T/J=2.1331$. The quantity $\Upsilon_{L}(z)/J$ is just the
helicity modulus determined for each layer separately.
The layered helicity  modulus is symmetric with respect to the middle layer 
where it reaches its maximum and decreases when the boundaries are approached. 
Although the helicity modulus $\Upsilon(T,H,L)/J$ is not the average of the
quantity $\Upsilon_{L}(z)/J$ over all layers, the curve in Fig. \ref{fig3}
is an approximation to the profile that the superfluid density develops 
in thin films. 
The basis for the standard scaling argument is the following. For large $H$ and
very close to the critical point where $\xi(T)$ is very large, the
``penetration'' depth $\lambda(T)$ of the superfluid density inside the film
is of the order of the correlation length. Thus, in the limit where all
other length scales are small compared to $H$ and $\xi$, if we plot $Y H$
versus $z/H$ (or $z/\xi$) we should find scaling.  However, for small
$H$ there is at least one length scale $D$ (which for $H \sim D$ needs
to taken into account) which has the following origin. The length scale
$D$ contains information on how fast $\Upsilon(z)$ rises from 
$\Upsilon(z=0)=0$.
Namely the $z$-derivative of $\Upsilon(z)$ is not universal, it depends on how
we have imposed the Dirichlet boundary conditions. There are many ways
to make the order parameter vanish at the boundary. It can be made to 
be zero when averaged over a boundary area $A=l\times l$.
In our case of staggered
boundary conditions $l=\sqrt{2}$. If we had chosen Dirichlet boundary
conditions where the order parameter is zero over an area with $l>\sqrt{2}$ we
would have found a slower rise of $\Upsilon(z)$ from its zero value at 
the boundary. If this initial ``faster rise'' of the superfluid density
is neglected, the rest of $\Upsilon(z)$ can be fit to $A \cosh(z/\xi)+B$ with
only one length scale, the correlation length $\xi$. Thus the curve
$Y(z)$ can be thought of as made of two contributions, and scaling 
at small values of $H$ can be obtained only if the film size
is extended. We can imagine that this thinner film of size $H$ is 
obtained from a thicker one by a process of forcing
the superfluid density to go to zero faster than its ``natural way''
by a ``speed'' dictated by the severity of the boundary conditions.

\begin{figure}[htp] 
 \centerline{\psfig {figure=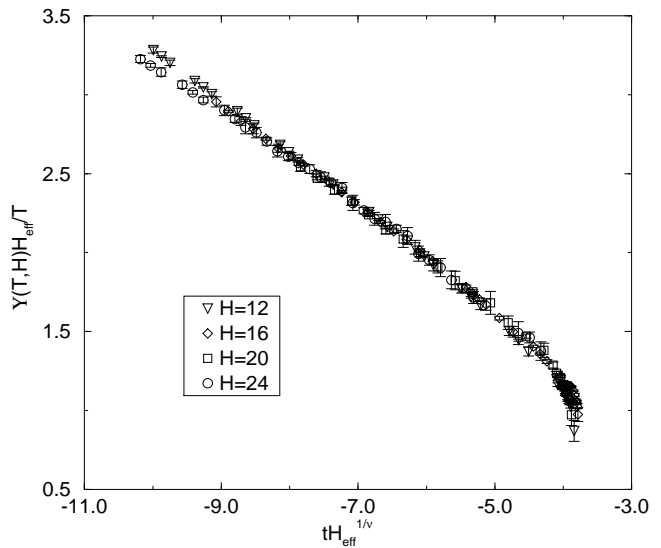,width=\mywidth}}
 \caption{\label{fig4}   $\Upsilon(T,H) H_{eff}/T$ as a function of 
$tH_{eff}^{1/\nu}$ for various thicknesses. $H_{eff}=H+5.79$ and $\nu=0.6705$.}
\end{figure}

The lack of scaling with the expected critical exponent $\nu=0.6705$ indicates 
that the critical temperatures $T_{c}^{2D}(H)$ do not satisfy Eq. (\ref{tch}).
Because of the argument given earlier about the profile of  the superfluid
density we may expect an
effective film thickness $H_{eff}$ to enter the scaling expressions
(\ref{tch}) and (\ref{tyh}). The simplest assumption is $H_{eff}=H+D$ where
$D$ is a constant. Indeed by replacing $H$ with $H_{eff}$ in these
equations for the film thicknesses $H=12,16,20$ we obtain
$x_{c}=-3.81(14)$ and $D=5.79(50)$ with $\nu=0.6705$. 
In Fig.\ref{fig4} we plot
$\Upsilon(T,H)H_{eff}/T$ as a function of $tH_{eff}^{1/\nu}$ for films
with $H=12,16,20,24$ where $\nu=0.6705$. The data for the helicity 
modulus collapse onto one universal curve.

We wish to test the assumption that the boundaries 
introduce an effective thickness into the scaling expression 
(\ref{tyh}) further.  Janke and Nather\cite{janna} 
studied the thickness dependence
of the Kosterlitz--Thouless transition temperature of the 
Villain model with open boundary conditions 
(interactions of the top and bottom layer only with the 
interior film layers). They found, however, that in order for scaling to
occur they needed to use a value for $\nu$ higher than
the value believed for the model. We replace  $H$ in Eq. (\ref{tch}) 
by the effective thickness $H_{eff}=H+D_{V}$.
Indeed, taking the expected value $\nu=0.6705$ we find  $D_{V}=1.05(2)$ and
$x_{c}=-1.62(2)$ and a good quality of fit. 
We can understand the increment $D$ as an effective scaling correction which 
renders the scaling relations (\ref{tch}) and 
(\ref{tyh}) valid even for very thin films. For large thicknesses
$H$ the increment $D$ can be neglected and we recover the
conventional scaling forms. This result 
means that the film thicknesses 
considered in Ref. \cite{janna} were still too small to extract
the expected value of the critical exponent $\nu$ from
the $H$--dependence of the critical temperature (\ref{tch})
without the help of an effective thickness $H+D_{V}$. 

\begin{figure}[htp] 
 \centerline{\psfig {figure=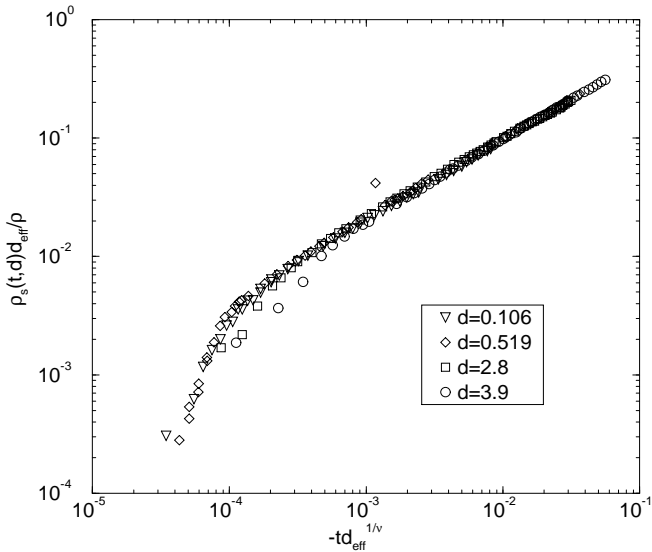,width=\mywidth}}
 \caption{\label{fig5} Scaling of the superfluid density data of Rhee et al.
\protect\cite{rhee} with the effective thickness $d_{eff}=d+0.145$.
$\nu=0.6705$ and all lengths are in $\mu m$.}
\end{figure}

In the experimental situation it is possible to imagine a similar
situation where a length scale $D$ emerges and corresponds to an
average defect distance on the substrate.
In Fig.\ref{fig5} we achieve approximate collapse of the data 
for the superfluid density $\rho_{s}$ for films of various 
thickness $d$ ($d$ is in $\mu m$) given in Refs. 
\cite{rhee} by plotting $\rho_{s}(t,d)d_{eff}/\rho$ 
versus $td_{eff}^{1/\nu}$ with
$\nu=0.6705$ and $d_{eff}=d+0.145$. We obtained the effective 
thickness by examining the reduced temperatures $t_{fs}(H)$ 
where finite--size effects set in. According to finite--size 
scaling theory $t_{fs}$ has to fulfill the 
relation $t_{fs}\propto d^{-1/\nu}$, thus in our case 
$t_{fs}\propto d_{eff}^{-1/\nu}$. 
The data points corresponding to the film with $d=3.9\mu m$
deviate from the universal curve; we attribute this to the 
anomalous behavior of these data. Namely, in general 
$|t_{fs}(d_{1})| > |t_{fs}(d_{2})|$ 
if $d_{1}<d_{2}$, but this is not the case for 
$d_{1}=2.8\mu m$ and $d_{2}=3.9\mu m$ 
(cf. Refs. \cite{rhee}). 
The scaling of the experimental data with the aid of this finite length
scale indicate that the average distance between defects will introduce
a new length scale from which one has to stay away in order to see
simple finite size scaling. Theoretically there are two different limits in
which simple scaling with $H$ should be expected to be valid. The clean
boundary limit, $\xi \sim H \ll D$, and the dirty boundary limit where
$D \ll \xi \sim H$.

In conclusion, we find that the results of our simulations for the
superfluid density on rather
small thickness films obey a scaling relation where an effective
thickness is introduced. Conventional scaling relations is expected to
be valid on films of much larger thickness. Applying the same idea of the
effective thickness on the experimental data, we found that the long
standing problem of lack of conventional scaling in the data of Rhee et
al\cite{rhee} can be resolved in a simple way without resorting to any
departure from scaling nor to using unrealistic values for $\nu$. Clearly,
more experiments with different substrates are desirable.
  
We wish to thank F. M. Gasparini for providing us with his data of the
superfluid density for films of various thickness and for clarifying 
discussions. This work was supported by the National Aeronautics and Space
Administration under grant no. NAGW-3326.

\end{document}